\documentclass[10pt,twocolumn,letterpaper]{article}

\usepackage{cvpr}
\usepackage{times}
\usepackage{epsfig}
\usepackage{graphicx}
\usepackage{amsmath}
\usepackage{amssymb}
\usepackage{authblk}
\usepackage{url}
\usepackage{multirow}

\usepackage[breaklinks=true,bookmarks=false]{hyperref}

\cvprfinalcopy 

\def\cvprPaperID{52} 

\begin{document}

\title{Hierarchical Regression Network for Spectral Reconstruction from RGB Images}

\author[1]{Yuzhi Zhao \thanks{Corresponding author: yzzhao2-c@my.cityu.edu.hk}}
\author[1]{Lai-Man Po}
\author[2]{Qiong Yan}
\author[2,3]{Wei Liu}
\author[1]{Tingyu Lin}
\affil[1]{City University of Hong Kong, Hong Kong SAR, China}
\affil[2]{SenseTime Research}
\affil[3]{Harbin Institute of Technology, China}

\renewcommand\Authands{ , } 

\maketitle
\pagestyle{empty}  
\thispagestyle{empty} 

\begin{abstract}

Capturing visual image with a hyperspectral camera has been successfully applied to many areas due to its narrow-band imaging technology. Hyperspectral reconstruction from RGB images denotes a reverse process of hyperspectral imaging by discovering an inverse response function. Current works mainly map RGB images directly to corresponding spectrum but do not consider context information explicitly. Moreover, the use of encoder-decoder pair in current algorithms leads to loss of information. To address these problems, we propose a 4-level Hierarchical Regression Network (HRNet) with PixelShuffle layer as inter-level interaction. Furthermore, we adopt a residual dense block to remove artifacts of real world RGB images and a residual global block to build attention mechanism for enlarging perceptive field. We evaluate proposed HRNet with other architectures and techniques by participating in NTIRE 2020 Challenge on Spectral Reconstruction from RGB Images. The HRNet is the winning method of track 2 - real world images and ranks 3rd on track 1 - clean images.

\end{abstract}

\section{Introduction}

Hyperspectral (HS) imaging technology refers to the spectral signature is densely sampled to many narrow bands. It combines imaging technology with spectral technology to detect the two-dimensional geometric space and one-dimensional spectral information of the target to obtain continuous, narrow-band images with high spectral resolution. Normally, most of the civil cameras capture only three primary colors. However, HS spectrometers can obtain the spectrum of each pixel in the scene and collect the information into a set of images. To visualize HS images, a response function is adopted to transform HS images into RGB format. Conversely, we can acquire HS images from the visible format by learning the inverse function. In this paper, we propose a general hierarchical regression network (HRNet) for spectral reconstruction from RGB images.

HS imaging technology has many advantages and particular characteristics. There have been many applications based on HS imaging technology, e.g, remote sensing technology \cite{melgani2004classification}, pedestrian detection \cite{hwang2015multispectral, liu2016multispectral}, food processing \cite{qin2013hyperspectral}, medical imaging \cite{andersson1987multispectral}. However, in recent years, the development of HS imaging has encountered a bottleneck since it mainly depends on spectrometers. The traditional spectrometers saves images with huge volume and need long operation time, which restricts HS imaging technology applied to portable platforms and high-speed moving scenes \cite{poli2012review}. Although researchers have continuously optimized the traditional pipeline \cite{cao2011prism, wang2016adaptive}, these hardware devices are still expensive and of high complexity. Thus, we present a low cost and automate approach only based on RGB cameras. To address the problem, we propose a HRNet that learns the process of RGB images to corresponding HS projections.


In general, spectral reconstruction is an ill-posed problem. Moreover, there is unknown noise in environment leading to degraded RGB images. However, there is dense correspondence between RGB images and HS images, making it possible to exploit the correlation from many RGB-HS pairs. Since the information of RGB image is much less than HS image, there may be many reasonable HS image combinations corresponding to a same RGB image. The algorithm needs to learn a reasonable mapping function that produces high-quality HS images. With the development of deep convolutional neural network (CNN), it is eligible to learn the blind mapping for spectral construction.

The previous methods \cite{shi2018hscnn+, koundinya20182d, stiebel2018reconstructing, can2018efficient, xiong2017hscnn} mainly utilize an auto-encoder structure with residual blocks \cite{he2016deep}. The network often performs convolution at low spatial resolution since the features are more compact and the computation is more efficient. However, as the network goes deeper, it fails to remain the original pixel information due to performing down-sampling by convolutions. To address this problem, we introduce a lossless and learnable sampling operator PixelShuffle \cite{shi2016real}. To further boost the quality of generated images, we propose a hierarchical architecture that extracts the features of different scales. At each level, the input is obtained by the reverse PixelShuffle (PixelUnShuffle) that no pixel is lost. Moreover, we propose to use residual dense block and residual global block in HRNet for removing artifacts and noise and modelling remote pixel correlation, respectively.

In general, there are three main contributions of this paper:

(1) We propose a HRNet that utilizes PixelUnShuffle and Pixelshuffle layers for downsampling and upsampling without information loss. We also propose residual dense block with residual global block to enlarge perceptive field and boost generation quality;

(2) We propose a 8-setting ensemble strategy to further enhance the generalization of HRNet;

(3) We evaluate proposed HRNet on NTIRE 2020 HS dataset. The HRNet is winning method of track 2 - real world images and ranks 3rd on track 1 - clean images.

\section{Related work}

\textbf{Hyperspectral image acquisition.} Conventional methods for hyperspectral image acquisition often adopt spectrograph with spatial scanning or spectral scanning technology. There are several types of scanner utilized for capturing images including pushbroom scanner, whiskbroom scanner, and band sequential scanner. They have been widely used to many applications such as detector, environmental monitoring and remote sensor for decades. For instance, pushbroom scanner and whiskbroom scanner are used for photogrammetric and remote sensing by satellite sensors \cite{poli2012review, breuer2000geometric}. However, those devices need to capture the spectral information of single points or bands separately, then scan the whole scene to get a fully HS image, which is difficult to capture scenes with moving objects. In addition, they are too large physically and not suitable for portable platforms. In order to address the problems, many kinds of non-scanning spectrometers have been developed to adapt the application of dynamic scenes \cite{descour1995computed, cao2011prism, wang2016adaptive}.

\textbf{Hyperspectral image reconstruction from RGB images.} Since the traditional methods for hyperspectral image acquisition are not portable or time-consuming for many applications, current methods attempt to reconstruct hyperspectral image from RGB image. By learning the mapping from RGB images to hyperspectral images on a big RGB-HS dataset, it is more convenient to obtain many HS images. Recent years have witnessed various studies including sparse coding and deep learning. In 2008, Parmar et al. \cite{parmar2008spatio} proposed a data sparsity expanding method to recover the spatial spectral data cube. Arad et al. \cite{arad2016sparse} first leveraged HS prior in order to create a sparse dictionary of HS signatures and their corresponding RGB projections. While Aeschbacher et al. \cite{aeschbacher2017defense} pushed the performance of Arad et al.'s method for better accuracy and runtime based on A+ framework \cite{timofte2014a+}.

Beyond the dataset provided by Arad et al. \cite{arad2016sparse}, many approaches proposed their own dataset. For instance, Yasuma et al. \cite{yasuma2010generalized} utilized a CCD camera (Apogee Alta U260) to captured 31-band multispectral images (400–700 nm, at 10 nm intervals) of several static scenes. Nguyen et al. \cite{nguyen2014training} captured a dataset by Specim's PFD-CL-65-V10E (400 nm to 1000 nm) spectral camera and there were total 64 images. Chakrabarti et al. \cite{chakrabarti2011statistics} explored a statistical model based on 55 HS images of indoor and outdoor scenes. With the improvement of the scale and resolution of natural HS dataset, the training of deep learning method becomes more feasible, a number of algorithms based on convolutional neural network were proposed \cite{koundinya20182d, stiebel2018reconstructing}. Simon et al. \cite{jegou2017one} proposed a fully convolutional densely connected “Tiramisu” network with one hundred layers for semantic segmentation. Galliani et al. \cite{galliani2017learned} enhanced it for spectral image super-resolution. Can et al. \cite{can2018efficient} improved it to avoid overfitting to the training data and obtain faster inference speed. Moreover, Xiong et al. \cite{xiong2017hscnn} proposed a unified HSCNN framework for hyperspectral recovery from both RGB and compressive measurements. To boost the performance, they developed a deep residual network named HSCNN-R, and another distinct architecture that replaces the residual block by the dense block with a novel fusion scheme, named HSCNN-D, collectively called HSCNN+ \cite{shi2018hscnn+}.

\textbf{Convolutional neural networks.} The convolutional neural networks have been successfully applied in many low-level vision tasks, e.g. colorization \cite{zhang2016colorful, isola2017image}, inpainting \cite{iizuka2017globally, yu2019free}, deblurring \cite{kupyn2018deblurgan}, denoising \cite{gu2019self, chen2018learning}, and demosaicking \cite{chen2018learning, zhao2019saliency}. Hyperspectral reconstruction, as one of low-level task, has gained great improvement of performance recently by deep convolutional neural networks. In order to facilitate convergence and extract features effectively, many well-known basic blocks are utilized in those frameworks such as residual block and dense block. He et al. \cite{he2016deep} proposed a residual network initially for image classification. It improves the accuracy obviously compared with traditional cascade convolutional structure. Then, the residual block has been widely used in image enhancement region for maintaining low-level features by the short connection. It was enhanced by densenet proposed by Huang et al. \cite{huang2017densely} to improve the feature fusion ability. Moreover, Hu et al. \cite{hu2018squeeze} strengthened them by a squeeze-and-excitation network including a feature attention mechanism. It was implemented by MLP layers for modelling connections of pixels in different spatial location. In general, our HRNet combines the advantages of above methods and provides a more effective and accurate solution for HS reconstruction.

\begin{figure}[htbp]
\centering
\includegraphics[width=\linewidth]{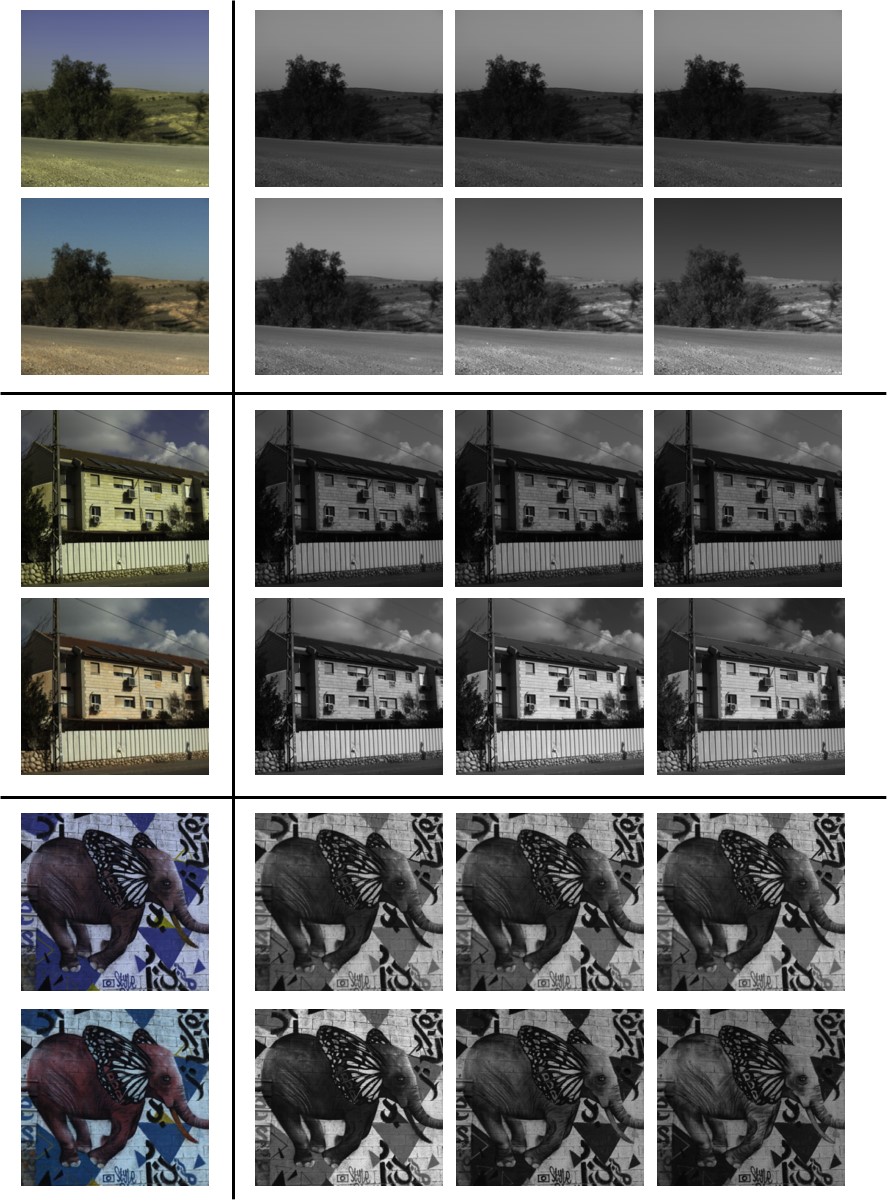}
\caption{Visualization of NTIRE 2020 HS dataset. For each group, from top to bottom and left to right, they represent clean RGB images, real world RGB images, HS images with 400 nm, 410 nm, 420 nm, 500 nm, 600 nm, and 700 nm channels, respectively.}
\label{sample}
\end{figure}

\begin{figure*}[t]
\centering
\includegraphics[width=\linewidth]{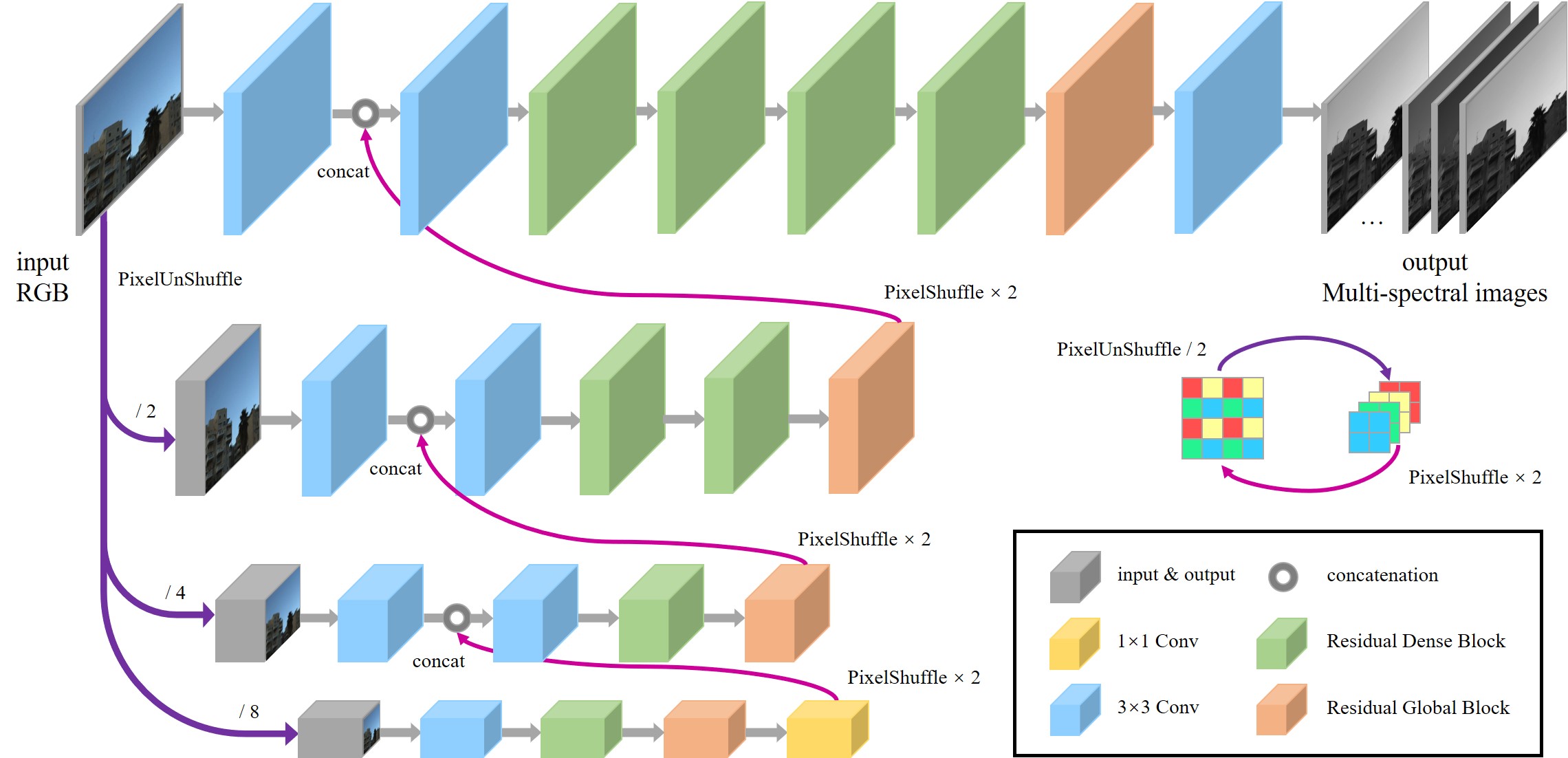}
\caption{Illustration of the architecture of HRNet. Please visit the project web page \url{https://github.com/zhaoyuzhi/Hierarchical-Regression-Network-for-Spectral-Reconstruction-from-RGB-Images} to try our codes and pre-trained models.}
\label{network}
\end{figure*}

\begin{figure}[htbp]
\centering
\includegraphics[width=\linewidth]{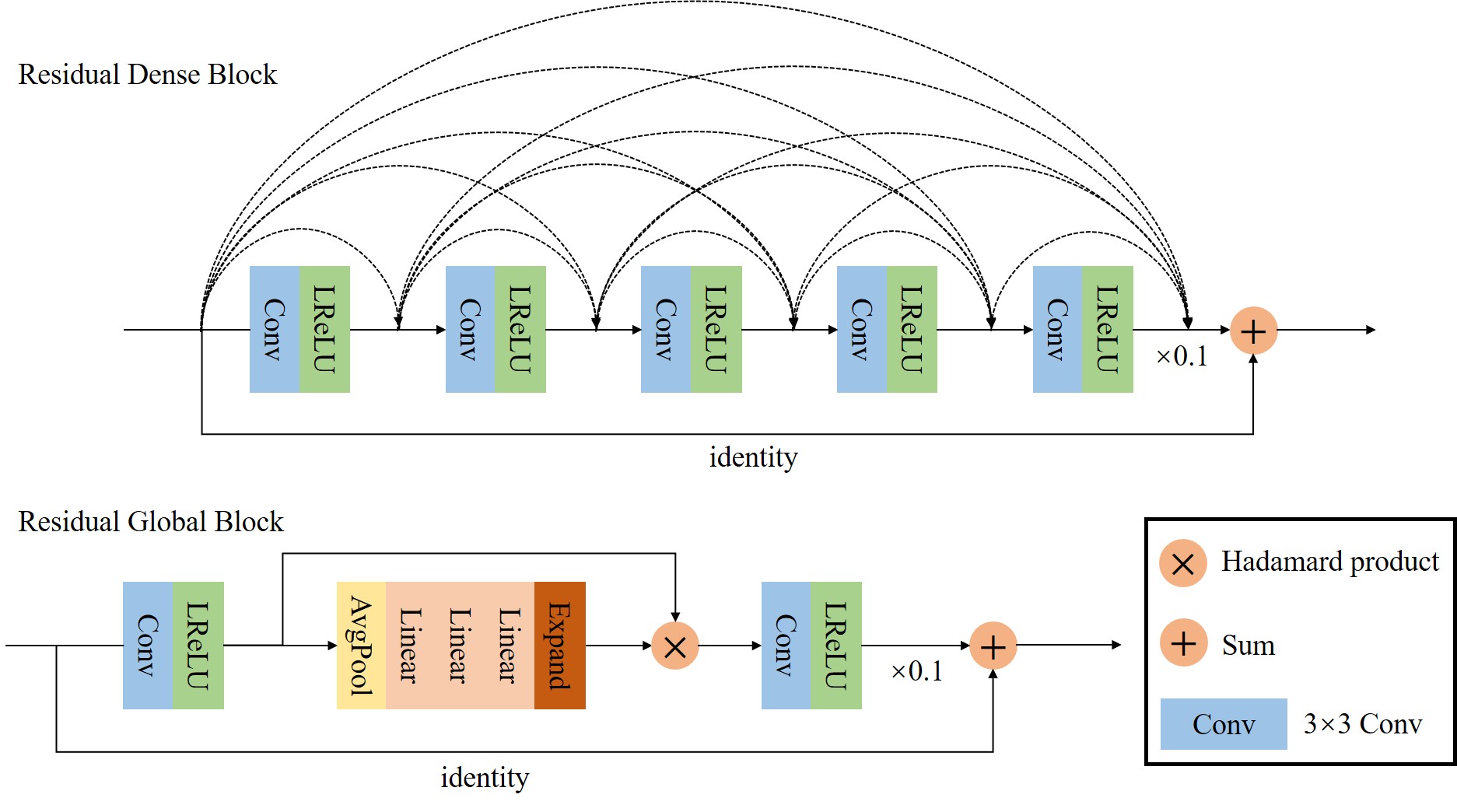}
\caption{Illustration of the architecture of residual dense block (ResDB) and residual global block (ResGB).}
\label{network2}
\end{figure}

\section{Methodology}

\subsection{Dataset}

We train our approach on the HS dataset provided by NTIRE Challenge 2020. This dataset consists of three parts: spectral images, clean RGB images (for track 1) and real world RGB images (for track 2). There are overall 450 RGB-HS pairs in training for both tracks involving different scenes. Each spectral image has the information of 31 bands in range of 400 nm to 700 nm. It is of $482 \times 512$ spatial resolution. To generate its corresponding RGB image, there is a fixed response function applied to HS bands. The rendering process can be defined as:

\begin{equation}
RGB = HS \times ResponseFunc.
\label{equ2}
\end{equation}

The RGB images and HS images include 3 and 31 channels, respectively. The $ResponseFunc$ maps each HS band to visible channel R, G, and B by 93 parameters. For clean RGB images, they are constructed by a known response function and saved as uncompressed format. However, the real world RGB images are acquired by unknown response function with additional blind noise and demosaicking operation. Some examples are illustrated in Figure \ref{sample} (e.g. 1st band approximately covers the 395-405 nm range).

\subsection{HRNet architecture}

Generally, we propose a 4-level network architecture for high-quality spectral reconstruction from RGB images, as shown in Figure \ref{network}. The PixelUnShuffle layers \cite{shi2016real} are utilized to downsample the input to each level without adding parameters. Therefore, the number of pixels of input is fixed while the spatial resolution decreases. Conversely, the learnable PixelShuffle layers are adopted to upsample feature maps and reduce channels for inter-level connection. The PixelShuffle only reshapes feature maps and does not introduces interpolation like bilinear upsampling. It allows the network to learn upsampling operation adaptively.

For each level, the process is decomposed to inter-level integration, artifacts reduction, and global feature extraction. For inter-level learning, the output features of subordinate level are pixel shuffled, then concatenated to current level, finally processed by an additional convolutional layer to unify channel number. In order to effectively reduce artifacts, we adopt residual dense block \cite{he2016deep, huang2017densely}, containing 5 dense-connected convolutional layers and a residual. Moreover, the residual global block \cite{he2016deep, hu2018squeeze} with short-cut connection of input is used to extract attention for every remote pixels by MLP layers.

Since the features are most compact in bottom level, there is a $1 \times 1$ convolutional layer attached to the last of bottom level in order to enhance tone mapping by weighting all channels. The two mid levels process features at different scales. Moreover, the top level uses the most blocks to effectively integrate features and reduce artifacts thus produce high-quality spectral images. The illustration of these blocks are in Figure \ref{network2}.

\subsection{Implementation details}

We only use L1 loss in the training process, which is a PSNR-oriented optimization for the system. The L1 loss is defined as:

\begin{equation}
L_{1} = \mathbb{E}[||G(x) - y||_1],
\end{equation}
where $x$ and $y$ are input and output, respectively. The $G(*)$ is the proposed HRNet. Note that, we utilize the local patches for efficient training. The input RGB image and output spectral images are cropped in same spatial region.

For network architecture, all the layers are LeakyReLU \cite{maas2013rectifier} activated except output layer. We do not use any normalization in HRNet to maintain the data distribution. The reflect padding is adopted for each convolutional layer in order to reduce border effect. The weights of VCGAN are initialized by Xavier algorithm \cite{glorot2010understanding}.

For training details, we use the entire NTIRE 2020 HS dataset (450 HS-RGB pairs for both tracks) at training. The whole HRNet is trained for 10000 epochs overall. The initial learning rate is $1 \times 10^{-4}$ and halved every 3000 epochs. For optimization, we use Adam optimizer with $\beta_1 = 0.5$ , $\beta_2 = 0.999$ and batch size equals to 8. The image pairs are randomly cropped to $256 \times 256$ region and normalized to range [0, 1]. All the experiments are implemented using 2 NVIDIA Titan Xp GPUs. It takes approximately 7 days for whole training process.

\subsection{Ensemble strategy}

Since the solution space of spectral reconstruction is often large, there may be multiple settings that achieve same performance on the training set. Therefore, a single network may lead to poor generalization performance since it tends to fall into local minima. However, we can minimize this risk by combining multiple network settings to enhance generalization and fuse the knowledge. In order to perform ensemble strategy, we use 4 other hyper-parameter settings and train HRNet from scratch for both tracks. These settings can be summarized as:

\begin{itemize}

\item Re-train the HRNet using baseline training setting.

\item Exchange the position of residual dense block and residual global block in HRNet, and use baseline training setting.

\item Train the network with different batch size (2 or 4) and keep other hyper-parameter settings, network architecture.

\item Train the network with different cropping patch size ($320 \times 320$ or $384 \times 384$) and keep other hyper-parameter settings, network architecture.

\end{itemize}

\begin{figure*}[htbp]
\centering
\includegraphics[width=\linewidth]{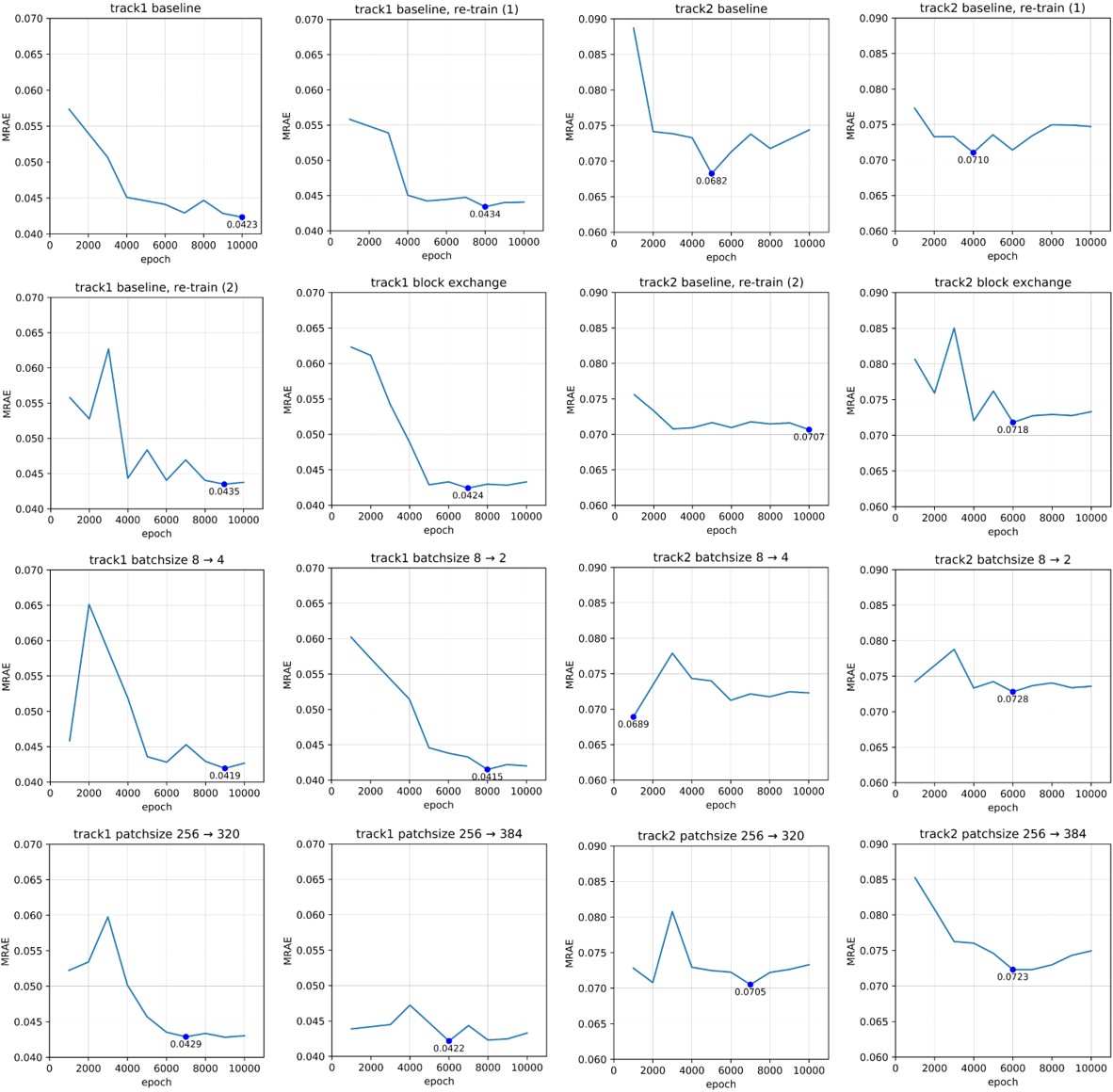}
\caption{The MRAE between ground truth spectral images and the generated images of different hyper-parameter settings for ensemble.}
\label{compare}
\end{figure*}

Therefore, there are 8 kinds of training methods. All the methods used for ensemble are trained for 10000 epochs. We record the MRAE (mean absolute value between all bands of generated spectral images $G(x)$ and ground truth $y$) every 1000 epochs, as shown in Table \ref{compare_table} and Figure \ref{compare}. Finally, we utilize the epoch with best MRAE value of 8 methods for computing average.

\begin{table}[t]
\begin{center}
\begin{tabular}{lcccc}
\hline
Setting & track 1 & track 2 \\
\hline
\hline
Baseline & 0.042328 & 0.068245 \\
Re-train baseline (1st) & 0.043408 & 0.071044 \\
Re-train baseline (2nd) & 0.043487 & 0.070668 \\
Exchange position of blocks & 0.042418 & 0.071798 \\
Change batch size 8 to 4 & 0.041936 & 0.071259 \\
Change batch size 8 to 2 & 0.041507 & 0.072797 \\
Change patch size 256 to 320 & 0.042810 & 0.070502 \\
Change patch size 256 to 384 & 0.042166 & 0.072313 \\
\hline
\hline
Ensemble & \textbf{0.039893} & \textbf{0.068081} \\
\hline
\end{tabular}
\end{center}
\caption{The best MRAE value of both tracks for HRNet settings used for ensemble.}
\label{compare_table}
\end{table}

\section{Experiment}

\subsection{Experimental settings}

We evaluate proposed HRNet by comparing with other network architectures and conducting ablation study on NTIRE 2020 HS dataset. For each track, there are 10 validation RGB images. The evaluation metrics are defined as:

\begin{itemize}

\item \textbf{MRAE.} It computes the pixel-wise disparity (mean absolute value) between all bands of generated spectral images $G(x)$ and ground truth $y$. It explicitly represents the construction quality of network. It is defined as:

\begin{equation}
MRAE = \frac{1}{N} \sum_{i=1}^{N}\frac{|G(x)^{i} - y^{i}|}{y^{i}},
\end{equation}
where $N$ denotes the overall pixels of spectral images.

\item \textbf{RMSE.} It computes the root mean
square error between the generated and ground truth spectral images with 31 bands. It is defined as:

\begin{equation}
RMSE = \sqrt{ \frac{1}{N} \sum_{i=1}^{N} (G(x)^{i} - y^{i})^2 } .
\end{equation}

\item \textbf{Back Projection MRAE (BPMRAE).} It evaluates the colorimetric accuracy of recovered RGB images from the generated and ground truth spectral images by a fixed camera response function. It is defined as:

\begin{equation}
BPMRAE = \frac{1}{N} \sum_{i=1}^{N}\frac{|(R \times G(x))^{i} - (R \times y)^{i}|}{y^{i}},
\end{equation}
where $R$ denotes the function $ResponseFunc$.

\end{itemize}

\subsection{Comparison with other architectures}


\begin{table}[t]
\begin{center}
\begin{tabular}{lcccc}
\hline
\multicolumn{2}{c}{Method} & U-Net & U-ResNet & HRNet \\
\hline
\hline
\multirow{2}{*}{MRAE} & track 1 & 0.047507 & 0.045242 & \textbf{0.042328} \\
 & track 2 & 0.074230 & 0.078892 & \textbf{0.068245} \\
\hline
\multirow{2}{*}{RMSE} & track 1 & 0.014154 & 0.013927 & \textbf{0.013537} \\
 & track 2 & 0.018647 & 0.020630 & \textbf{0.017859} \\
\hline
\multirow{2}{*}{BPMRAE} & track 1 & 0.007926 & 0.007171 & \textbf{0.006064} \\
 & track 2 & 0.044966 & 0.055876 & \textbf{0.042105} \\
\hline
\end{tabular}
\end{center}
\caption{The quantitative comparison results of different architectures and HRNet on NTIRE 2020 HS validation set.}
\label{comparison}
\end{table}

We utilize two common network architectures for comparison: U-Net \cite{ronneberger2015u} and U-ResNet \cite{ronneberger2015u, he2016deep}. Both of them have been widely used in many previous low-level tasks \cite{isola2017image, iizuka2017globally, yu2019free, kupyn2018deblurgan, chen2018learning, zhao2019saliency}. The first convolutional layer and last convolutional layer utilize $7 \times 7$ convolution without changing spatial resolution. The training scheme for all methods are same. Other details are concluded as: (1) U-Net. The encoder layers perform convolution with stride of 2. The spatial resolution of bottom feature map equals to $1 \times 1$. There are short concatenations between each encoder layer and decoder layer with same resolution; (2) U-ResNet. The total number of encoder layers and decoder layers are half of U-Net. Instead, there are 4 residual blocks attached to the last layer of encoder. The concatenations are reserved.

\begin{figure*}[t]
\centering
\includegraphics[width=\linewidth]{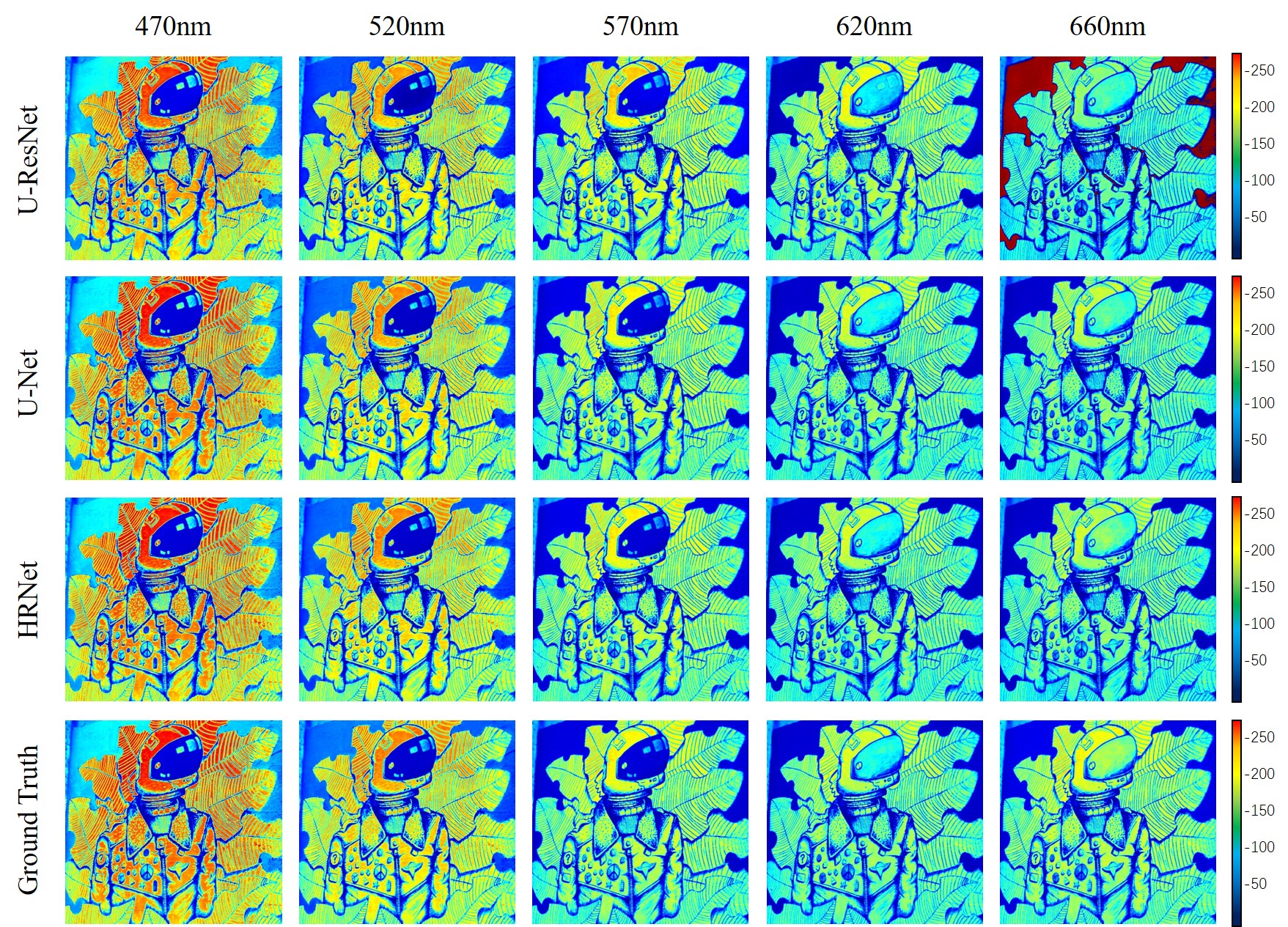}
\caption{Visualization of generated results from U-ResNet, U-Net, and proposed HRNet on NTIRE 2020 HS validation set track 1.}
\label{v1}
\end{figure*}

\begin{figure*}[t]
\centering
\includegraphics[width=\linewidth]{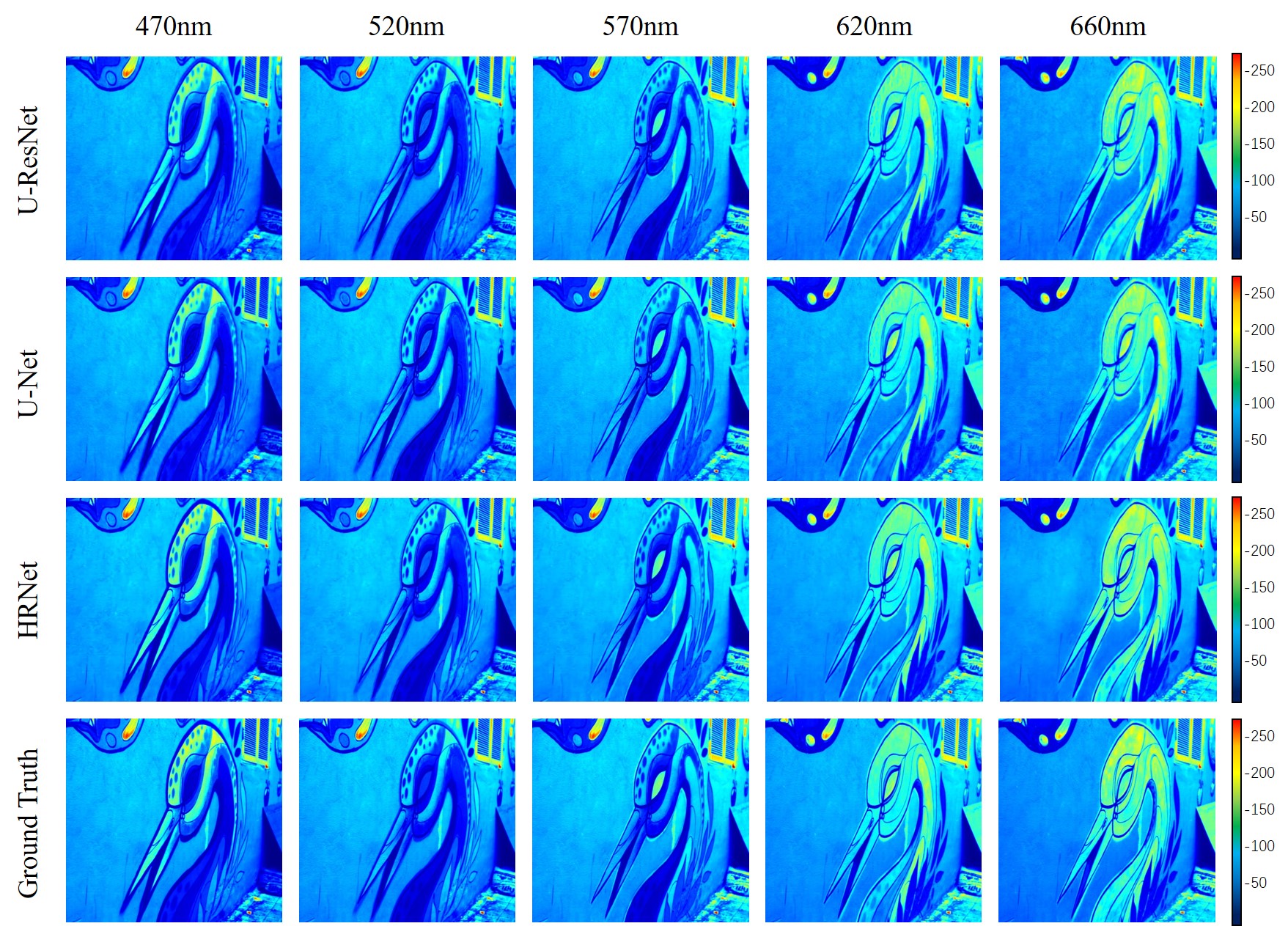}
\caption{Visualization of generated results from U-ResNet, U-Net, and proposed HRNet on NTIRE 2020 HS validation set track 2.}
\label{v2}
\end{figure*}

We train both networks using same hyper-parameters of HRNet until convergence. There is no ensemble strategy used. We generate the reconstructed spectral images using the best epoch of them. The results are summarized in Table \ref{comparison}. We also visualize each method in Figure \ref{v1} and \ref{v2} by pseudo-color map. The first three rows show the data distribution of 3 methods and last row indicates ground truth. We recommend readers to compare textures of background.

There are two reasons that proposed HRNet outperforms other two methods. The first is that HRNet utilizes PixelShuffle to connect each level. Traditional nearest or bilinear upsampling will introduce redundancy information to features, which is unnecessary for feature extraction. However, by the combination of PixelUnShuffle and PixelShuffle, HRNet could process high-level features more efficiently. The second is that HRNet adopts two residual-based blocks, which facilitate convergence and assist each level to exploit different scales of features. Moreover, the blocks with residual learning helps remove artifacts. The residual global block enhances context information since it models correlation for every two pixels.

\begin{table}[t]
\begin{center}
\begin{tabular}{lcccc}
\hline
Method & w/o ResDB & w/o ResGB & w/o both & HRNet \\
\hline
\hline
MRAE & 0.042448 & 0.042565 & 0.048033 & \textbf{0.042328} \\
RMSE & 0.014216 & 0.014092 & 0.015740 & \textbf{0.013537} \\
BPMRAE & 0.009507 & 0.007669 & 0.015502 & \textbf{0.006064} \\
\hline
\end{tabular}
\end{center}
\caption{The comparison results of ablation study on NTIRE 2020 HS validation set track 1 - clean images.}
\label{ablation}
\end{table}

\begin{table}[t]
\begin{center}
\begin{tabular}{lcccc}
\hline
Method & HRNet ($\frac{1}{2}$) & HRNet ($\frac{1}{4}$) & HRNet ($\frac{1}{8}$) \\
\hline
\hline
MACs (G) & 46.413 & 12.017 & 3.212 \\
Params (Mb) & 8.185 & 2.176 & 0.6088 \\
Weights (Mb) & 32.006 & 8.532 & 2.410 \\
MRAE & 0.042457 & 0.046424 & 0.048443 \\
RMSE & 0.015147 & 0.015459 & 0.015659 \\
BPMRAE & 0.006886 & 0.007806 & 0.009891 \\
\hline
\end{tabular}
\end{center}
\caption{The comparison results of compressed HRNet model (the number of channels decreased to $\frac{1}{2}$, $\frac{1}{4}$, and $\frac{1}{8}$ of the original) on NTIRE 2020 HS validation set track 1 - clean images.}
\label{para}
\end{table}

\begin{table*}[t]
\begin{center}
\begin{tabular}{lccc}
\hline
Team & MRAE & Runtime / Image (seconds) & Compute Platform \\
\hline
\hline
Deep-imagelab & 0.03010476377 & 0.56 & 2$\times$NVIDIA 2080Ti \\
\hline
ppplang & 0.03075687151 & 16 & NVIDIA 1080Ti \\
\hline
\textbf{HRNet} & \textbf{0.03231183605} & 3.748 & 2$\times$NVIDIA Titan Xp \\
\hline
ZHU\_ zy & 0.03475963089 & ~1 & \emph{Unknown} \\
\hline
sunnyvick & 0.03516495956 & 0.7 & Tesla K80 12GB \\
\hline
\end{tabular}
\end{center}
\caption{The final testing results of NTIRE 2020 Spectral Reconstruction from RGB Images Challenge track 1 - clean images.}
\label{track1}
\end{table*}

\begin{table*}[t]
\begin{center}
\begin{tabular}{lccc}
\hline
Team & MRAE & Runtime / Image (seconds) & Compute Platform \\
\hline
\hline
\textbf{HRNet} & \textbf{0.06200744887} & 3.748 & 2$\times$NVIDIA Titan Xp \\
\hline
ppplang & 0.06212710705 & 16 & NVIDIA 1080Ti \\
\hline
Deep-imagelab & 0.06216655487 & 0.56 & 2$\times$NVIDIA 2080Ti \\
\hline
PARASITE & 0.06514769779 & ~30 & NVIDIA Titan Xp \\
\hline
Tasti & 0.06732598306 & \emph{Unknown} & NVIDIA 2080Ti \\
\hline
\end{tabular}
\end{center}
\caption{The final testing results of NTIRE 2020 Spectral Reconstruction from RGB Images Challenge track 2 - real world images.}
\label{track2}
\end{table*}

\subsection{Ablation study}

In order to demonstrate the effectiveness of both residual dense block (ResDB) and residual global block (ResGB), we replace them by plain convolution layers with similar FLOPs. The results in track 1 - clean images is shown in Table \ref{ablation}. The baseline of HRNet is shown in Table \ref{compare_table}, which has better performance comparing with all ablation settings. If we delete all ResDB or ResGB in HRNet, the MRAE decreases the most, which demonstrates the combination of both blocks is significant for spectral reconstruction.

We conduct another experiment that shrinks the HRNet model size by decreasing channels of each convolutional layer to half, one fourth, and one eighth of original numbers. It will compress model size greatly by sacrificing pixel fidelity. To better compare these settings, we conclude the multiply–accumulate operation (MACs), total network parameters (Params), model size saved on machine (Weights) and 3 quantitative metrics results in Table \ref{para}. The MACs, Params, and Weights of baseline HRNet are 182.347 Gb, 31.705 Mb, and 123.879 Mb, respectively. Users can choose high-quality HRNet to obtain high pixel fidelity of  spectral images (MRAE $=$ 0.042328) or high-efficiency HRNet with small size (Weights $=$ 2.410 Mb).

\subsection{Testing result on NTIRE 2020 challenge}

The proposed HRNet ranks 3rd and 1st on track 1 and track 2, respectively, of NTIRE 2020 Spectral Reconstruction from RGB Images Challenge \cite{NTIREChallenge}. The comparison results on testing set are summarized in Table \ref{track1} and \ref{track2}. Moreover, the HRNet has better performance on track 2 since it adopts two effective blocks for removing artifacts while utilizes learnable PixelShuffle upsampling operator. The ensemble strategy works obviously on both tracks that improves the MRAE from 0.042328 to 0.039893 since it avoids the HRNet to fall into local minima. In conclusion, both HRNet architecture and ensemble strategy contribute to spectral reconstruction performance.

\section{Conclusion}

In this paper, we presented a 4-level HRNet for automatically generating spectrum from RGB images. For each level, it adopts both residual dense block and residual global block for effectively extracting features. While the PixelShuffle is utilized for inter-level connection. Then, we proposed a novel 8-setting ensemble strategy to further enhance the quality of predicted spectral images. Finally, we validated the HRNet outperforms the well-known low-level vision frameworks such as U-Net and U-ResNet on NTIRE 2020 HS dataset. Furthermore, we presented 3 types of compressed HRNets and analyzed their reconstruction performance and computing efficiency. The proposed HRNet is the winning method of track 2 - real world images and ranks 3rd on track 1 - clean images.

{\small
\bibliographystyle{ieee_fullname}
\bibliography{egbib}
}

\end{document}